\documentclass[a4paper]{article}
\usepackage{graphicx}
\usepackage{twocolceurws}
\usepackage[ruled,vlined]{algorithm2e}
\usepackage{adjustbox}
\usepackage{caption} 
\usepackage{hyperref}
\usepackage{tcolorbox}
\usepackage{listings}
\usepackage[utf8]{inputenc}

\title{XPath Agent: An Efficient XPath Programming Agent Based on LLM for Web Crawler}

\author{
Yu Li \\ \\ lijingyu68@gmail.com
\and
Bryce Wang\textsuperscript{*} \\ Stanford University  \\ brycewang2018@gmail.com
\and
Xinyu Luan \\ \\ xinyuluan0320@gmail.com
}

\institution{}

\begin{document}
\maketitle

\begin{abstract}
  We present XPath Agent, a production-ready XPath programming agent specifically designed for web crawling and web GUI testing. A key feature of XPath Agent is its ability to automatically generate XPath queries from a set of sampled web pages using a single natural language query. To demonstrate its effectiveness, we benchmark XPath Agent against a state-of-the-art XPath programming agent across a range of web crawling tasks. Our results show that XPath Agent achieves comparable performance metrics while significantly reducing token usage and improving clock-time efficiency. The well-designed two-stage pipeline allows for seamless integration into existing web crawling or web GUI testing workflows, thereby saving time and effort in manual XPath query development. The source code for XPath Agent is available at \url{https://github.com/eavae/feilian}.
\end{abstract}

\section{Introduction}

Web scraping\cite{khder2021web} automates data extraction from websites, vital for modern fields like Business Intelligence. It excels in gathering structured data from unstructured sources like HTML, especially when machine-readable formats are unavailable. Web scraping provides real-time data, such as pricing from retail sites, and can offer insights into illicit activities like darknet drug markets.

The advent of HTML5\cite{TABARES2021101529} has introduced significant complexities to automated web scraping. These complexities stem from the enhanced capabilities and dynamic nature of HTML5, which require more sophisticated methods to accurately extract and interpret data. To address these challenges, researchers have developed a variety of tools, such as Selenium\cite{selenium}. Which offers a set of application programming interfaces (APIs) to automate web browsers, enabling the extraction of data from web pages. However, program Selenium to extract data from web pages is a time-consuming and error-prone process. The most crucial part is how to locate the target information on the web page. XPath queries provide a solution to this problem by allowing users to navigate the HTML structure of a web page and extract specific elements. But, programming XPath queries is a challenging task, especially for non-technical users.

The development of Large Language Models (LLM) has emerged as a promising avenue. LLMs, with their advanced natural language processing capabilities, offer a new paradigm for understanding and interacting with web content. AutoWebGLM\cite{lai2024autowebglmlargelanguagemodelbased} demonstrated significant advancements in addressing the complexities of real-world web navigation tasks, particularly in simplifying HTML data representation to enhancing it's capability. By leveraging reinforcement learning and rejection sampling, AutoWebGLM enhanced its ability to comprehend webpages, execute browser operations, and efficiently decompose tasks.

Instead of one time task execution, AutoScraper\cite{huang2024autoscraperprogressiveunderstandingweb} adopt a simplified technique which only involves text content of webpages. By focusing on the hierarchical structure of HTML and traversing the webpage, it construct a final XPath using generated action sequences. Such XPath is generalizable and can be applied to multiple webpages. Which significantly reduce the time required when execution.

But, the above approaches are not efficient in generating XPath queries. We introduced a more effective approach to generate XPath queries using LLMs which could simply integrate into existing web crawling or web GUI testing workflows. 

\subsection{Motivation}

We assuming there are 3 core reasons why LLMs are not efficient in generating XPath queries. Firstly, LLMs are not designed to generate XPath queries. Secondly, web pages are lengthy and complex, full of task unrelated information. Those information distract LLMs from generating the correct XPath queries. Thirdly, LLMs are context limited. A good XPath query should be generalizable across different web pages. However, LLMs can only generate XPath queries based on the context they have seen. So, a shorter and more task-related context is more likely to generate a better XPath query.

Based on the above insights, we propose a novel approach to generate XPath queries using LLMs. We aim to reduce the number of steps required to generate a well-crafted XPath query, reduce the computational overhead, and improve the generalizability of the XPath queries generated by LLMs.

In order to increase the efficiency of XPath query generation, we also employed LangGraph. Which is a graph-based tool set which we can define the whole pipeline in a graph-based manner and execute it in parallel. Which significantly reduce the time to generate XPath queries.

\subsection{Our Contributions}

In summary, our contributions are as follows:

\begin{enumerate}
  \item We designed a two stage pipeline, which we can employ a weaker LLM to extract target information. And a stronger LLM to program XPath.
  \item We proposed a simple way to prune the web page, which can reduce the complexity of the web page and make the LLM focus on the target information.
  \item We discovered that extracted cue texts from 1st stage significantly improve the performance of the 2nd stage.
  \item We benchmarked our approach against a state-of-the-art same purpose agent across a suite of web crawling tasks. Our findings reveal that our approach excels in F1 score with minimal compromise on accuracy, while significantly reducing token usage and increase clock-time efficiency.
\end{enumerate}
\section{Related Work}
Information extraction (IE) serves as a cornerstone of natural language processing, transforming unstructured text into structured entities, relations, and events. Traditional IE approaches often relied on task-specific models, which demanded substantial annotated corpora and computational resources. The advent of large language models (LLMs), particularly GPT-based architectures, has revolutionized this field. LLMs excel in zero-shot and few-shot learning due to pretraining on vast text corpora, enabling efficient and flexible IE for diverse and complex applications.

\subsection{Web Information Extraction}

Web information extraction (WebIE) has evolved alongside the increasing complexity of web layouts and dynamic content. Traditional rule-based methods, effective for static pages, struggle with modern web structures. LLMs and advanced frameworks have emerged as solutions, leveraging schema-based, multimodal, and generative techniques to unify tasks like Named Entity Recognition (NER), Relation Extraction (RE), and Event Extraction (EE) through prompt engineering and fine-tuning \cite{xu2024largelanguagemodelsgenerative}.

Recent works emphasize the importance of modeling structural context in webpages. WebFormer \cite{wang2022webformerwebpagetransformerstructure} employs graph attention mechanisms to capture local and global dependencies within the DOM tree, achieving state-of-the-art results on the SWDE benchmark. Similarly, FreeDOM \cite{DBLP:journals/corr/abs-2010-10755} combines node-level encoding with pairwise relational features, enabling robust generalization across unseen websites without visual rendering. Simplified DOM representations, such as those used in SimpDOM \cite{zhou2021simplifieddomtreestransferable}, enhance feature transferability across verticals, achieving superior cross-domain performance while avoiding reliance on computationally expensive visual features.

Generative approaches further demonstrate the adaptability of LLMs in WebIE. ChatIE \cite{wei2024chatie} reformulates extraction tasks into multi-turn dialogues, improving zero-shot capabilities. Frameworks like NeuScraper \cite{ahluwalia2024leveraginglargelanguagemodels} integrate neural models with HTML structures to enhance data quality, while MIND2WEB \cite{deng2023mind2webgeneralistagentweb} extends these innovations with a dataset and framework for training generalist agents capable of performing diverse tasks across real-world websites.

On the multimodal front, models like GoLLIE \cite{sainz2024gollieannotationguidelinesimprove} and MoEEF \cite{yang2024hypertextentityextractionwebpage} integrate visual and textual data to address domain-specific challenges. GoLLIE aligns annotation guidelines with code generation for better extractions, while MoEEF incorporates hypertext features to boost entity recognition. Vision-based LLMs, such as GPT-4 Vision\cite{fellman-etal-2024-future}, highlight the growing importance of multimodal techniques for processing visually encoded web data.

\subsection{LLMs and XPaths for Web Information Extraction}

XPath, a language for querying hierarchical document structures, plays a vital role in bridging structured representation with computational modeling. MarkupLM\cite{DBLP:journals/corr/abs-2110-08518} demonstrates how XPath captures the genealogy of DOM nodes, seamlessly integrating textual and markup features for tasks such as Masked Markup Language Modeling (MMLM) and Node Relation Prediction (NRP). These techniques enable dynamic adaptation to complex web layouts, surpassing the limitations of 2D positional embeddings.

XPath generation further extends WebIE capabilities. Tools like TREEX \cite{10.1145/3018661.3018740} and AUTOSCRAPER\cite{huang2024autoscraperprogressiveunderstandingweb} employ decision tree learning and progressive HTML traversal to generate reusable extractors, balancing precision and recall for diverse applications such as price comparison and product aggregation. By leveraging the interpretive capabilities of LLMs, these methods scale to diverse web environments, underscoring the synergy between structured document analysis and scalable information extraction.

\subsection{Summary}

The evolution of WebIE reflects the integration of structural, multimodal, and generative techniques into scalable systems. LLMs and frameworks leveraging DOM structures, multimodal data, and XPath queries have significantly advanced this field, enabling robust and adaptable solutions for dynamic web environments. These advancements collectively highlight the transformative potential of combining pre-trained models, structural embeddings, and domain-specific optimizations in tackling modern WebIE challenges. Future directions should focus on enhancing generalization across unseen web domains and improving multimodal fusion for visually rich webpages efficiently.

\begin{figure*}[h]
  \centering
  \includegraphics[width=1\textwidth]{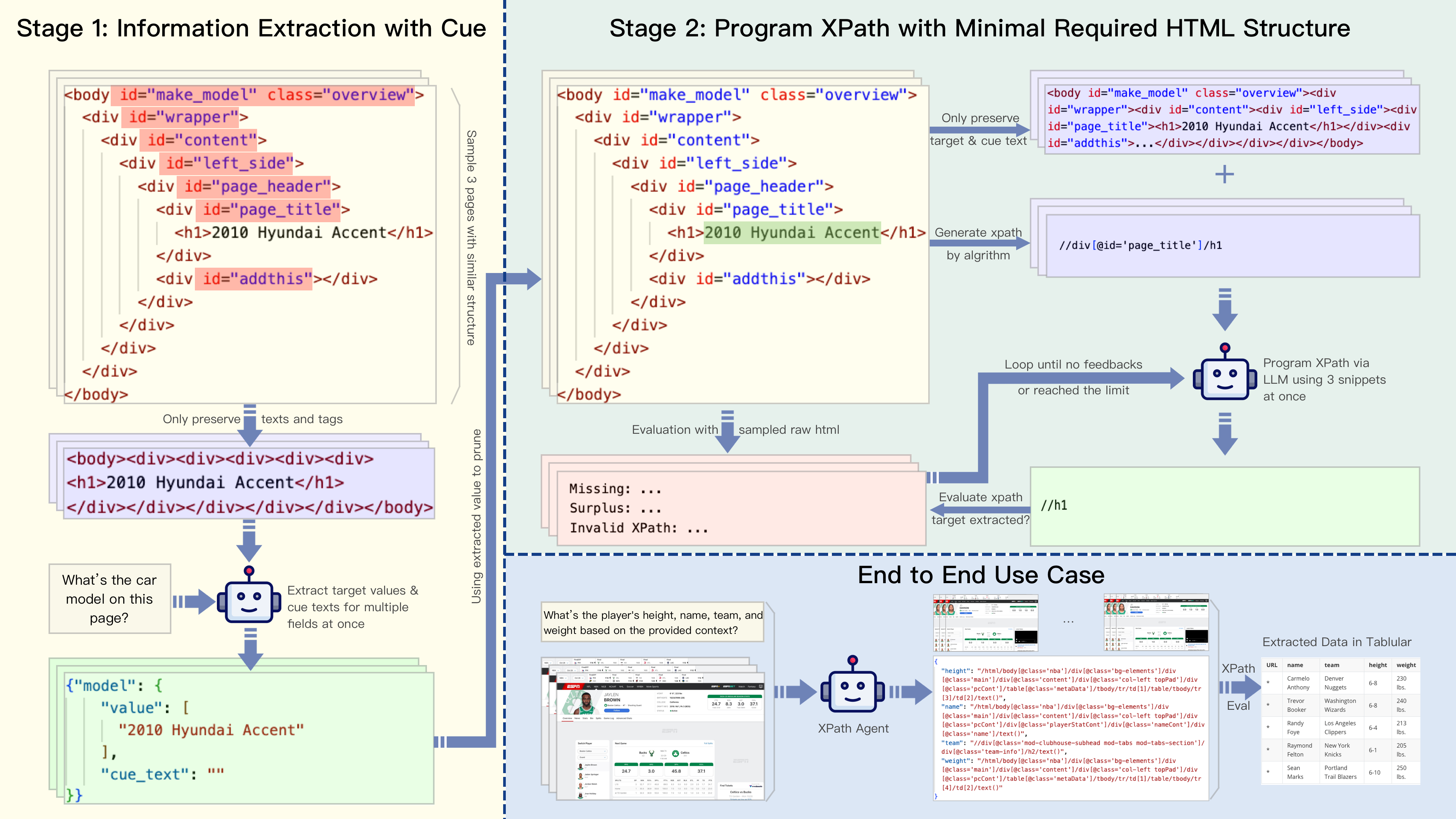}
  \caption{XPath Agent of two stages pipeline. The first stage is Information Extraction, which extracts target information and cue text from sanitized web pages (the red are sanitized). The second stage is XPath Programming, which generates XPath queries based on condensed html (the greens are target nodes) and generated XPath.}
  \label{fig:workflow}
  \vspace{20pt} 
\end{figure*}

\section{Methodology}

In this section, we present the methodology of our approach, which consists of two stages: Information Extraction (IE) and XPath Programming. The IE stage extracts target information from sanitized web pages, while the XPath Programming stage generates XPath queries based on the condensed html by extracted information. The whole process based on seeded web pages, which is 3 in our implementation. Figure \ref{fig:workflow} illustrates the two-stage pipeline of XPath Agent. For each stage, we provide a detailed description of the process and the algorithms used.

\subsection{Information Extraction with Cue Text}

The Information Extraction (IE) stage aims to extract target information. Not like traditional IE, we discovered 2 key insights. Firstly, we prompt LLM not only extract questioned information but also cue texts. Secondly, we sanitized the web page to reduce the complexity of the web page and make the LLM focus on the target information with contextual semantic.

\begin{figure}[h]
  \centering
  \includegraphics[width=0.5\textwidth]{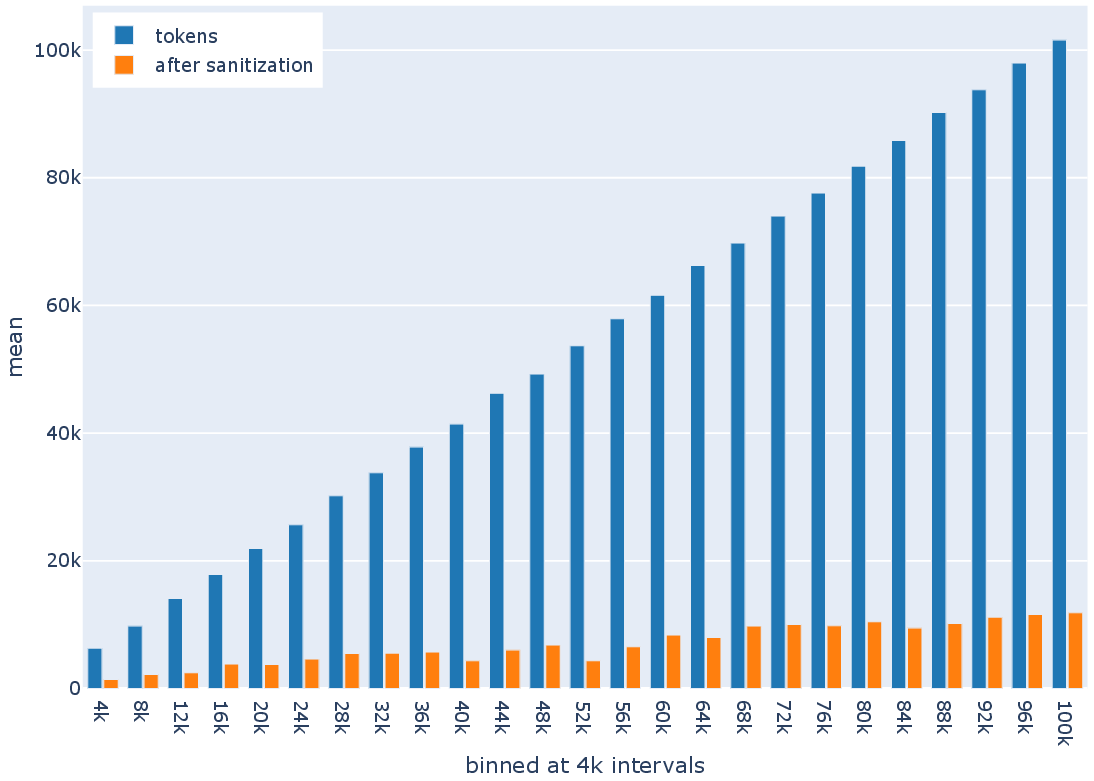}
  \caption{Token Stats Analysis with Algorithm 1. As page size grow, the size after sanitization increased slowly (sampled 128 pages for each category from SWDE dataset, around 10k pages totally).}
  \label{fig:ie_token_stats}
\end{figure}

Cue texts are the indicative texts that signals the upcoming target information. For example, for "price: \$100.00", the cue text is "price:". Those texts are important in some case, especially when no way or hard to directly programming XPath queries to extract target information "\$100.00". In such case, treat "price:" as an anchor, using XPath's ancestor, sibling, or descendant axis traverse to the target information is the only way. In order to let the context still be condensed, we prompt LLM to response cue texts simultaneously.

Sanitizing web page is a process to remove unnecessary information from a page. In HTML, the most meaningful parts are texts and tags. The texts are the target information we want to extract, and the tags are the structure of the web page which tells the relationship between texts especially the priority of which answer is more likely to be the target information. The purpose of sanitizing the web page is to reduce the complexity of the web page and make the LLM focus on the target information. We designed an algorithm to sanitize the web page, which is shown in Algorithm \ref{alg:sanitizer}. We also employed minify-html\cite{minifyhtml} to further reduce the size of the web page.

The algorithm 1 traverse the HTML tree in a depth-first manner. It removes the invisible or empty nodes, and all attributes. It's efficient and can be easily implemented in any programming language. In our sampled web pages, it can help us reduce the size of the web page to 10\%~20\% on average.

In our implementation, we prompt LLM to extract all information at once on a single page with JSON format. So, multiple fields or multiple values for a single field might be extracted. We treat all extracted values are relevant and passing them to the next stage. The prompt for the Information Extraction stage is shown in Section \ref{sec:ie_prompt}.

\begin{algorithm}
  \SetAlgoLined
  \caption{IE HTML Sanitizer}
  \label{alg:sanitizer}
  \KwIn{Root node of HTML tree $root\_node$}
  \KwOut{Sanitized HTML tree}
  
  $left\_stack \gets [root\_node]$\;
  $right\_stack \gets []$\;
  
  \While{$left\_stack$ is not empty}{
      $node \gets left\_stack.pop()$\;
      $right\_stack.append(node)$\;
      $left\_stack \gets left\_stack + list(node.iterchildren())$\;
  }
  
  \While{$right\_stack$ is not empty}{
      $node \gets right\_stack.pop()$\;
      \If{$is\_invisiable\_or\_no\_text(node)$}{
          $node.getparent().remove(node)$\;
      }
      \Else{
          $node.remove\_attributes()$\;
      }
  }
\end{algorithm}

\subsection{Program XPath}

Program XPath queries is a process to generate XPath queries based on the condensed html by extracted information. In order to let LLM have more context to program a robust XPath query, we condensed the html by the extracted information and prompt with 3 seeded web pages at once. The algorithm is shown in Algorithm \ref{alg:condenser}.

The condenser based on the extracted information, which is the target information and cue texts. A distance function is used to identify the most relevant nodes so that we can keep them in the condensed html. During the condense process, we keep the target nodes and replace other nodes' children with "...".

\subsection{Static XPath Generation}

Program a robust XPath query is a challenging task. Which requires to balance between rigidity and flexibility. The rigidity means the XPath query should strictly follow the specific structure of the web page. The flexibility means the XPath query should be pruned to be generalizable across different web pages. In our early experiments, we discovered that the XPath query generated by LLM is not sticky to the structure of the web page. So, we designed a static XPath generation algorithm to guide LLMs.

The static XPath generation algorithm propagate from the target node to root node. It generates the XPath query in a bottom-up manner. Unlike naive XPath generation algorithm. We add more attributes (in our case, we include class and id) to the XPath query. It makes the XPath query richer.

\subsection{Conversational XPath Evaluator}

LLM are lacks of the environment to evaluate, correct or improve the XPath queries. So, we designed a evaluator to evaluate LLM generated XPath.

The XPath evaluator is a function, which execute the XPath query on seeded web pages and feedback the result to LLM. The result include 3 parts: what are missing, what are redundant, and correctness of the XPath query. The LLM can based on feedback to improve the XPath query.

In our implementation, we limited the number of feedback loop to 3. At the end of the loop, we take the best XPath query based on evaluation result as the final result. The prompt for the Program XPath stage is shown in Section \ref{sec:program_xpath_prompt}.

\begin{algorithm}
  \SetAlgoLined
  \caption{HTML Condenser}
  \label{alg:condenser}
  \KwIn{
      $root$: HTML root node\;
      $target\_texts$: List of target texts to keep\;
      $d$: Distance function between two texts\;
  }
  \KwOut{
      $root$: Condensed HTML root node\;
  }
  
  $target\_texts \gets []$\;
  $distances \gets \{\}$\;
  $eles \gets \{\}$\;
  
  \ForEach{$ele, text$ in \texttt{iter\_with\_text}($root$)}{
      \ForEach{$target\_text$ in $target\_texts$}{
          $distance \gets d(text, target\_text)$\;
          \If{$distance < distances[text]$}{
              $distances[text] \gets distance$\;
              $eles[text] \gets [\texttt{get\_xpath}($ele$)]$\;
          }
          \ElseIf{$distance == distances[text]$}{
              $eles[text].\texttt{append}(\texttt{get\_xpath}($ele$))$\;
          }
      }
  }

  $targets \gets \texttt{concat}(\texttt{values}(eles))$\;
  \ForEach{$xpath, ele$ in \texttt{iter\_with\_xpath}($root$)}{
    \If{\texttt{is\_outside}($xpath$, $targets$)}{
      \texttt{remove\_children}(ele)\;
      \texttt{replace\_text\_to}(ele, "...")\;
    }
  }

\end{algorithm}

\section{Experiments }

\subsection{Experimental Setup}

\subsubsection{Models}
We use DeepSeek and ChatGPT-4.0 as the primary large language models in our experiments.

\subsubsection{Dataset}
We use the SWDE\cite{abdin2024phi3technicalreporthighly} (Structured Web Data Extraction) dataset, which includes 90 of websites across 9 domains, in total 20414 web pages.

\subsubsection{Experimental Parameters}
We use DeepSeek-Chat as the main model for our XPath Agent. The Number of Seeds is 3 initial seeds are provided to guide query generation with a  
sample size of 32 web pages are sampled per task to evaluate the model’s adaptability and generalizability. 

\subsection{Evaluation Metrics}
For evaluation, we employ the metrics of precision, recall, F1 score, and accuracy. we utilized a set matching method to calculate the metrics 
for multi-label classification tasks where the labels are unordered. 
This involves comparing each predicted label set with the corresponding ground truth label set while disregarding the order of the labels. 
First, Both the ground truth and predictions are converted into sets to ignore any order. Accuracy is then defined such that a prediction is 
counted as correct if the predicted set exactly matches the ground truth set. We count correctly predicted labels as True Positives. 
Incorrectly predicted labels that are not in ground truth are classified as False Positives, while the labels that should have been in predicted set 
but were missed are designated as False Negatives. Additionally, since our data does not contain blank answers (with only a very few exceptions that have been removed), 
our cases do not have true negatives. Finally, precision, recall, F1 score, and accuracy are calculated by the following formula.
\begin{center}
$presision=\frac{Ture Positives}{Ture Positives + False Positives}$\\ [0.3 cm]
$recall=\frac{Ture Positives}{True Positives + False Negatives}$\\ [0.3 cm]
$F1=\frac{2 \times precision \times recall}{precision + recall}$\\ [0.3 cm]
$accuracy=\frac{True Positives}{True Positives + False Positives + False Negatives}$
\end{center}

\section{Results and Analysis}
\subsection{Statistical Analysis}

\begin{table*}[h]
  \begin{center}
    \begin{tabular}{|c|c|c|c|c|}
    \hline
    \textbf{Model} & \textbf{Accuracy} & \textbf{Precision} & \textbf{Recall} & \textbf{F1} \\ \hline
    DeepSeek V2.5 & 0.5794 & 0.6764 & 0.8017 & 0.7337 \\ 
    GPT 4o mini & 0.6613 &0.8056 &0.7868 &0.7961 \\
    GPT 4o      &\textbf{0.7413} &\textbf{0.8561} &0.8467 &\textbf{0.8514} \\
    Claude 3.5  & 0.6191 & 0.6916 & \textbf{0.8551} & 0.7647 \\ \hline
    \end{tabular}
  \end{center}
  \caption{Experimental Results}
\end{table*}
    
DeepSeek is strong in precision, making it good at avoiding irrelevant data. GPT 4o is highly effective at capturing a wide range of relevant content, ensuring that few important elements are missed. Claude 3.0 strikes a balanced approach, effectively combining both precision and recall for solid overall performance of F1 score. Claude 3.5 stands out as the most balanced and effective model, excelling across all key areas of accuracy, precision, recall, and overall performance. It provides the best mix of identifying relevant data while minimizing errors.

\subsection{Comparative Analysis} 
The AutoCrawler\cite{huang2024autoscraperprogressiveunderstandingweb} framework focuses on generating web crawlers for extracting specific information from semi-structured HTML. 
It is designed with a two-phase approach: the first phase uses a progressive generation framework that leverages the hierarchical 
structure of HTML pages, while the second phase employs a synthesis framework that improves crawler performance by learning from 
multiple web pages. 

In comparison to XPath Agent, AutoCrawler presents a different approach, emphasizing a sequence of XPath actions rather than just the extraction of XPath from snapshots. This difference may influence performance in various metrics, such as F1 score. 
AutoCrawler's focus on refining action sequences based on learning from past errors might offer advantages in terms of robustness 
and adaptability to dynamic web structures. However, your XPath Agent, by isolating XPath extraction tasks, might achieve greater 
precision in structured environments where precise element identification is crucial.

\subsection{Error Analysis} TODO

\section{Conclusion }

Third level headings must be flush left, initial caps and bold.
One line space before the third level heading and $1/2$ line
space after the third level heading.

\paragraph{Fourth Level Heading}

Fourth level headings must be flush left, initial caps and roman type.
One line space before the fourth level heading and $1/2$ line
space after the fourth level heading.

\bibliography{refs}{}

\clearpage
\onecolumn

\appendix

\centerline{\Large\bfseries Appendix}

\section{Information Extraction Prompt}
\label{sec:ie_prompt}

The prompt for the Information Extraction stage in the following format:

\begin{tcolorbox}
  \begin{lstlisting}[language={}, basicstyle=\ttfamily, columns=fullflexible, breaklines=true]
Extract the information and cues from the given context that are required by the question, and present the results in JSON format. When presenting the results, ensure character-level consistency with the extracted text, do not make any modifications. The given context may be incomplete or may not have the answer, so, the final JSON conclusion must not include any fields that have not been mentioned.

When there are multiple similar expressions, prioritize them according to the following rules (from high to low):
1. The label of the target text is more important in HTML semantics.
2. The target text is completely within a tag, rather than within a sentence or paragraph.
3. The target text is closer to other fields to be extracted.
4. If these expressions can complement each other, please extract them all.

Cue Text (cue_text, from high to low):
1. Cue Text: In HTML, the indicative text that signals the upcoming extraction of the target text, such as `Phone number` or `Address:`.
2. When there is no cue text, use an empty string.

# Question:
{{ query }}

# Context:
```html
{{ context }}
```

# Answer Format (ignore the format requirements in the `Question`, strictly follow the answer format of cue_text and value):
Thought: ...(Your thoughts, about fields mentioned in the context and their cues)...
Conclusion: ...(Strictly follow the JSON example format to answer: {{ json_example }})...

# Your Answer:
\end{lstlisting}
\end{tcolorbox}

\section{Program XPath Prompt}
\label{sec:program_xpath_prompt}

The system prompt for the Program XPath stage in the following format:

\begin{tcolorbox}
  \begin{lstlisting}[language={}, basicstyle=\ttfamily, columns=fullflexible, breaklines=true]
You are a pro software engineer, your task is reading the HTML code that user sent, and then response **one** Xpath (wrapped in JSON) that can recognize the element in the HTML to extract `target value`. 

Here're some hints:
1. Do not output the xpath with exact value or element appears in the HTML.
2. Reference to the `target value` and the generated the xpath (if exists) to get more context.
3. When using text predication, always using `contains(., 'value')` instead of `text()='value'`.
4. If the target xpath ends with `text()[n]`, where n is not 1, please do not ignore it.
5. If cue text exist, using cue text and cue xpath to compose a new xpath might be a better idea.
6. String functions are allowed, such as `starts-with()`, `ends-with()`, `substring-before()`, `substring-after()`. Use it in caution, since it can only be used on `text()` node.

Please always response in the following Json format:
{
  "thought": "", # a brief thought of how to confirm the value and generate the xpath
  "xpath": "" # a workable xpath to extract the value in the HTML
}    
  \end{lstlisting}
\end{tcolorbox}

The feedback prompt for the Program XPath stage in the following format:

\begin{tcolorbox}
  \begin{lstlisting}[language={}, basicstyle=\ttfamily, columns=fullflexible, breaklines=true]
Following the feedbacks to refine the xpath you provided:

1. Extend the xpath to include the missing information if `Missing`.
2. Restrict the xpath to exclude the irrelevant information if `Surplus`.
3. Correct the xpath grammer if `Invalid`.
4. Response same xpath if no better solution.

{% for feedback in feedbacks %}
#### Evaluated on Fragment {{ feedback.id }}:
Extracted (JSON encoded): `{{ feedback.extracted | tojson }}`
Feedback Message: `{{ feedback.message }}`
{%- endfor %}    
  \end{lstlisting}
\end{tcolorbox}

\end{document}